\documentstyle[aaspp4,epsfig]{article}

\newcommand{\be}{\begin{equation}}
\newcommand{\ee}{\end{equation}}
\newcommand{\bea}{\begin{eqnarray}}
\newcommand{\eea}{\end{eqnarray}}
\newcommand{\nn}{\nonumber}

\newcommand{\appgeq}{\stackrel{>}{\sim}}
\newcommand{\appleq}{\stackrel{<}{\sim}}

\newcommand{\Gz}{\Gamma_z}
\newcommand{\Gphi}{\Gamma_\phi}
\newcommand{\Bz}{B_z}
\newcommand{\Bphi}{B_\phi}

\newcommand{\Cperp}{C_{\perp,j}}
\newcommand{\Cpar}{C_{\parallel,j}}
\newcommand{\Cpol}{C_{pol,j}}
\newcommand{\Cext}{C_{ext,j}}
\newcommand{\Cp}{C_{+,j}}
\newcommand{\Bnu}{B_\nu(T_j)} 
\newcommand{\alphaj}{\alpha_j}
\newcommand{\meanalpha}{\langle\alpha\rangle}

\newcommand{\Bzs}{B_{z,S}}
\newcommand{\Bphis}{B_{\phi,S}}
\newcommand{\ratio}{\Bzs/\Bphis}

\lefthead{Fiege \& Pudritz}
\righthead{Polarized Sub-Millimetre Emission from Filamentary Molecular Clouds}

\begin{document}

\title{Polarized Sub-Millimetre Emission from Filamentary Molecular Clouds}
\author{Jason D. Fiege$^{1,2}$ and Ralph E. Pudritz$^2$}
\affil{
$^1$Canadian Institute for Theoretical Astrophysics, \\
McLennan Labs, University of Toronto \\
60 St. George Street, Toronto, Ontario, M5S 3H8 \\
$^2$Dept. of Physics and Astronomy, McMaster University, \\
1280 Main St. W., Hamilton, Ontario, L8S 4M1\\
email: fiege@cita.utoronto.ca, pudritz@physics.mcmaster.ca}

\begin{abstract}
We model the sub-millimetre polarization patterns that are expected for filamentary clouds
that are threaded by helical magnetic fields.  We previously developed a three parameter
model of such clouds (Fiege \& Pudritz 2000a), which are described by a concentration parameter $C$, and two flux 
to mass ratios $\Gz$ and $\Gphi$ to specify the mass loading of the poloidal and toroidal field lines respectively. 
Our models provide a simple and purely geometric explanation for the well-known ``polarization hole'' effect, in 
which the sub-millimetre polarization percentage decreases toward the regions of peak intensity.
This occurs because of a cancellation between contributions to the polarization from the 
``backbone'' of poloidal flux along the filament's axis
and its surrounding envelope, which is dominated by the toroidal field component.
A systematic exploration of our parameter space allows us to classify the polarization patterns due to filaments aligned approximately 
perpendicular to the plane of the sky into three basic types.  The polarization vectors are parallel to 
filaments when $\ratio\appleq 0.1$, where $\Bzs$ and $\Bphis$ are respectively the poloidal and toroidal magnetic
field components at the outer surface of the filament.
The polarization vectors are perpendicular to filaments when $\ratio\appgeq 0.33$.  
Intermediate cases result in polarization patterns that contain $90^\circ$ flips in the orientation of the polarization vectors.
The flips are symmetric about the central axis for filaments oriented parallel to the plane of the sky, but more complicated
asymmetric patters result from filaments that are inclined at some angle.

\end{abstract}


\section{Introduction}
\label{sec:intro}

Molecular clouds are highly filamentary structures that are supported against rapid, global collapse by
a combination of ordered magnetic fields as well as by non-thermal, hydromagnetic turbulence of some kind.
However, determining the structure of the field has proven difficult.  Direct Zeeman measurements have been
limited to a few points per cloud rather than full-scale maps, and rarely trace the highest 
density gas (Crutcher et al. 1993, 1996, 1999).  Polarization maps in optical and near-infrared extinction fail to 
reveal the structure of the field in the dense gas, perhaps as a result of the poor polarizing power of 
grains at these wavelengths in regions of high optical depth (Goodman et al. 1995).  

Observations at far-infrared and sub-millimetre wavelengths have demonstrated that the thermal
emission from dust grains is often partially polarized (eg. Schleuning 1998,
Matthews \& Wilson 2000, Hildebrand et al. 1999; see also review by Weintraub, Goodman, \& Akeson 2000), 
which is usually attributed to the partial alignment of the emitting grains by the magnetic field.  
The mechanisms that are responsible for grain alignment are not entirely understood, but it is 
most often argued that dissipative processes gradually cause rapidly spinning grains to relax to a state
in which grains are, on average, aligned perpendicular to the magnetic field (cf. Lazarian, Goodman, \& Myers 1997).
The possibility now exists, therefore, of using completely sampled JCMT polarimetry maps to place
strong constraints on the structure of the magnetic fields in molecular clouds.
 
Current observations reveal the presence of well-ordered magnetic fields within molecular clouds and their 
clumps and cores.  However, the results are quite puzzling in that they often show
that the percent polarization $P$ {\em decreases} 
toward the highest density and most heavily obscured regions (eg. Schleuning
1998, Matthews \& Wilson 2000; see also Weintraub et al. 2000 and references therein).
This so-called ``polarization hole'' effect appears to be an almost universal feature of polarization maps
at these wavelengths, although its precise origin has been a mystery.  Nevertheless, two possible
explanations have been proposed.  Depolarization at high optical depths has often been attributed to the 
poor alignment of grains in dense regions (Goodman et al. 1995).  The second possibility is that the effect might be due
to disordered, sub-resolution scale variations in the orientation of aligned grains in dense regions, whose combined
polarized emission would tend to mostly cancel (see review by Weintraub et al. 2000, for example).  
Recent high resolution interferometric polarization maps at millimetre wavelengths by Rao et al. (1998) have revealed
strong, ordered polarization patterns {\em within} the polarization hole found by Schleuning.  The position angle
of the polarization vectors changes rapidly on the plane of the sky, which might be due to 
magnetic fields that are twisted on small scales.

The KL core is embedded in the ``integral-shaped'' filament of Orion A (See Johnstone \& Bally 1999).
There is some evidence that the polarization hole in the Orion filament is a global feature of the filament, and not
localized to the KL region.
Part of this filament has recently been mapped in 850 ${\mu m}$ polarization by Matthews and Wilson (2000),
who find that the polarization vectors are remarkably ordered and generally aligned with the filament.
They also find evidence for a decrease in the polarization percentage along the symmetry axis of the
filament (See Matthews \& Wilson 2000, Figure 2).

This paper demonstrates that a very simple explanation of the
polarization hole effect is possible, namely that it arises as a consequence of a more general
configuration of the magnetic field within filamentary clouds than has
previously been contemplated by many authors.  We show in this work that this effect can be 
rather naturally explained as arising from grain alignment by large-scale, ordered, helical magnetic fields
threading filamentary molecular clouds.
The Matthews and Wilson (2000) data suggest that the magnetic field threading the integral-shaped filament is probably
very well-ordered.  It is therefore conceivable that the polarization hole in this region
might be due to a large scale, orderly twist in the magnetic field, rather than disordered magnetic fields on small
scales.  In addition, Vall\'ee and Bastien (1999) compared the observed 760$\mu m$ polarization patterns 
of a number of sources with the qualitative patterns expected for several field geometries.  For the subset of 
maps which show an ordered magnetic field, they argued that helical fields are most consistent with the data. 

We present here, a very simple model of the polarized thermal emission of aligned grains to calculate the
polarization patterns that would be expected from our helically magnetized
filamentary cloud models (Fiege \& Pudritz 2000a,b; hereafter FP1 and FP2).  
Polarimetric maps are difficult to interpret because very different three-dimensional field geometries
can give rise to qualitatively similar polarization patterns (see the discussion in 
Sections \ref{sec:transverse} and \ref{sec:results} for example).
This degeneracy is compounded by the high degree of uncertainty regarding the composition and
distribution of grains, their polarization cross sections, and their alignment efficiencies.  However, 
we show that under the simplest possible set of assumptions, helical field models can produce polarization 
patterns that are not unlike the observed patterns.  Our most important result is that a large
fraction of our models are depolarized along the symmetry axis of the filamentary cloud, as a
result of the twisted field lines.  Thus helical magnetic fields might provide a very simple, and
purely geometric, explanation for the polarization hole phenomenon.

\section{Filamentary Clouds and Helical Magnetic Fields}
\label{sec:review}

We first summarize the basic properties of our model of magnetized filamentary clouds (see FP1 and FP2 for details).
Our models assume that filamentary clouds are self-gravitating, magnetized, truncated by an external pressure, 
and in a state of radial quasi-equilibrium between gravity, magnetic stresses, and internal pressure gradients.
The ISM provides the external pressure for many filaments, but some are embedded in molecular clouds (ie. the 
integral-shaped filament of Orion A).  We assume that the magnetic field is cylindrically symmetric,
with both a poloidal and toroidal component, so that the field is helical in general.
Note that the toroidal field component must vanish on the filament axis, where the poloidal field component
is strongest.  As one moves out in radius, the winding of the field on each subsequent cylinder increases as the
toroidal field strength relative to the poloidal field increases.   At large radii, the toroidal
field decreases, but less rapidly than the poloidal field.

We constructed a three parameter family of models for
helically magnetized filamentary clouds, which we constrained with an
observational data set collected from the literature.  We refer the reader to FP1 for the precise definition
and a full discussion of the model parameters.  Briefly, however, our model involves a concentration parameter $C$,
which defines the radius of pressure truncation, and two flux to mass ratios $\Gz$ and $\Gphi$, for the poloidal and
toroidal field components respectively, which are assumed to be
constant within each filament.  The models are characterized by a core radius $r_0$
defined as $\sigma (4\pi G\rho_c)^{-1/2}$, where
$\sigma$ is the one dimensional velocity dispersion of the gas (assumed to be constant),
$G$ is the gravitational constant, and $\rho_c$ is the central density along
the axis of the filament.  The core radius characterizes the approximate radial scale at which the density structure
changes from being nearly constant inside of $\sim r_0$, to an approximately $r^{-2}$ profile
in the less dense outer envelope.  The fact that we obtain an
approximately $r^{-2}$ profile is significant because three recent observations (Alves et al. 1998, Lada, Alves, \& Lada 
1999, and Johnstone \& Bally 1999) all find density gradients that are consistent with an $r^{-2}$ profile.
Our models are therefore in good agreement with the existing data on the radial density structure of filamentary clouds.

\section{A Simple Model for Sub-Millimetre Polarization by Aligned Dust Grains}
\label{sec:model}
The model that we adopt for the sub-millimetre polarization arising from our filamentary cloud models 
most closely resembles the analysis by Wardle and K\"{o}nigl (1990, hereafter WK90) for the polarized emission from the Galactic centre disk.
The main difference is that our approach allows us to estimate the magnitude of the polarization percentage,
whereas WK90 worked with normalized quantities.  Another difference is that we include the possibility of 
several grain species, with the assumption that the emitting grains are a uniformly mixed population
with the number density of each individual species proportional to the gas density.  
In addition, we require each grain species to be of uniform temperature and aligned to the same extent 
throughout the cloud.  

Our assumptions regarding the constancy of grain properties and their alignment efficiencies throughout
molecular clouds is undoubtedly a simplification of the true state of the grains in some clouds.  As an example of the
debate concerning grain properties, it has often been argued that the grains within dense regions
of molecular clouds do not efficiently polarize starlight in extinction because they are either nearly spherical
or poorly aligned at high optical depths (Goodman et al. 1995).  There are several reasons to think that this might
be the case.  For instance, grain growth due to coagulation tends to produce slightly larger, and possibly more 
spherical grains in regions of higher density (Vrba et al. 1993), thus decreasing their contribution to the 
polarized emission.  This effect might be compounded by the growth of ice mantles at high optical depths 
(Eiroa \& Hodapp 1989, Goodman et al. 1995).
The nature and efficiency of grain alignment mechanisms at high optical depths is perhaps
even less certain than the grain properties themselves.  The alignment might be poor because all efficient alignment mechanisms require
supra-thermal rotational velocities, and neither the Purcell (1979) mechanism 
nor the radiative alignment mechanism (Draine \& Weingartner 1996, 1997)
are likely to be efficient at high optical depths. 
\footnote{The radiative mechanism might be important within $A_v\approx 2$ of the surface of the cloud or 
any embedded sources (Draine \& Weingartner 1996).}
Nevertheless, simple, quantitative estimates of the alignment efficiency and distribution of grain shapes for
various physical conditions are presently lacking.  It is for this reason that we mainly consider the 
simplest possible model of constant magnetic alignment throughout the cloud.  
We note however that our models can be readily extended to explicitly include
radiative alignment.

\subsection{Analysis}
\label{sec:analysis}
Grains result in polarized thermal emission if they are aspherical and partially aligned, most probably by a magnetic field.  We consider the contribution to the polarized emission
from several grain species $j$, whose absorption cross sections
parallel and perpendicular to the symmetry axis are given by
$C_{\parallel,j}$ and $C_{\perp,j}$.  The thermal emission from these grains is polarized if
$C_{\parallel,j}$ and $C_{\perp,j}$ are not equal.
It is convenient to define the polarization cross section 
\be
\Cpol=\left\{
\begin{array}{ll} 
	\Cperp - \Cpar 	& \mbox{(oblate grains)}\\
	1/2(\Cpar - \Cperp) 	& \mbox{(prolate grains)}
\end{array}
\right.
\ee
(Lee \& Draine 1985, hereafter LD85; WK90).

To calculate the Stokes parameters $I$, $Q$, and $U$, which completely specify the intensity of the 
emission and its state of linear polarization,
we sum the contributions arising from each species of dust within the cloud.  We assume that 
species $j$ has temperature $T_j$, polarization cross section $\Cpol$, and number 
density $n_j=c_j \rho$, where $c_j$ is a constant and $\rho$ is the total mass density of the cloud.
We also assume that the mean alignment of the spin angular momentum with the magnetic field is given by 
\be
cos^2 \gamma_j=\left<\frac{ ({\bf B \cdot J})^2 } { B^2 J^2 }\right>.
\ee

A full analysis of the polarized emission applicable at all wavelengths would include the effects of
absorption and scattering.  However, both of these effects can be neglected if we restrict our analysis to
sub-millimetre wavelengths longer than about $100\mu m$ since the thermal emission is almost always optically thin
at these long wavelengths (Hildebrand 1983) and scattering is completely insignificant (Novak et al. 1989).
For example, the peak optical depth of even the dense KL core in Orion is only $\approx 0.25$ at $100\mu m$ 
(Novak et al. 1989, Schleuning 1998), and falls off with increasing wavelength as $\lambda^{-\beta}$, where 
$\beta \approx 1.5-2$ (Hildebrand 1983, Andr\'e, Ward-Thompson, \& Barsony 2000).  Therefore, it is appropriate to treat the 
sub-millimetre polarization as arising purely from emission in most circumstances.  

Following DL85, we define the polarization reduction factor $\Phi_j$ for dust grain 
species $j$ as 
\be
\Phi_j=R_j F_j \cos^2 \gamma,
\ee
where $R_j$ is the Rayleigh polarization reduction factor due to imperfect grain alignment, 
$F_j$ is the polarization reduction due to the turbulent component of the magnetic field, and
$\gamma$ is the angle between the plane of the sky and the local direction of the
magnetic field.  We assume that $R_j$ and $F_j$ are constants for each grain species.
The combination of variables $C_{pol,j} \Phi_j$ plays the role of the effective polarization 
cross-section, so that the contributions to the Stokes parameters $Q$
and $U$ from grains of species $j$ are given by the following:
\bea
Q_j &=& \Cpol R_j F_j c_j \Bnu c_j q \label{eq:Q} \\
U_j &=& \Cpol R_j F_j c_j \Bnu c_j u \label{eq:U},
\eea
where $B_\nu$ is the Planck function of the grain at temperature $T_j$, 
$ds$ is a distance element along the line of sight.  The quantities $q$ and $u$
are integrals along the line of sight $\hat{s}$, which depend on the density structure and the 
geometry of the magnetic field, but not the properties of the grains.  We define the $\hat{x}-\hat{y}$ 
plane as being parallel to the plane of the sky, with $\hat{x}$ pointing West and $\hat{y}$ pointing
North.  The angle $\psi$ is defined as the angle (counter-clockwise) between $\hat{y}$
and the projection of the magnetic field in the plane of the sky.  
Then, $q$ and $u$ are respectively given by  
\bea
q &=& \int \rho \cos 2\psi \cos^2 \gamma ~ds \label{eq:q} \\
u &=& \int \rho \sin 2\psi \cos^2 \gamma ~ds \label{eq:u}.
\eea
The angle $\psi$ varies as a function of $s$ along the line of sight for 3-dimensional field geometries
where the local direction of the plane-of-sky field component varies along the line of sight.
Note that $q$ and $u$ are identical in form to the ``relative Stokes parameters''
defined by WK90.  

The polarization angle $\chi$ is determined by solving the equations
\bea
\cos 2\chi=\frac{q}{\sqrt{q^2+u^2}}, \label{eq:angle1} \\
\sin 2\chi=\frac{u}{\sqrt{q^2+u^2}}  \label{eq:angle2}
\eea
Since equations \ref{eq:angle1} and \ref{eq:angle2} are multi-valued, we restrict $\chi$ to be on the range $[0,\pi)$.
The special case where $u=0$ and $q$ remains finite is very important to the analysis that follows.
For this case, the polarization angle is given by $\chi=0$ when $q>0$ and $\chi=\pi/2$ when $q<0$.

We proceed to calculate the intensity of the emission, which requires the 
extinction cross-section from equation 3.14 of LD85:
\be
\Cext=\Cp\left[1-\alphaj \left(\frac{\cos^2\gamma}{2}-\frac{1}{3} \right) \right],
\ee
where $\Cp$ and $\alphaj$ are defined as 
\bea
\Cp &=& \frac{2\Cperp+\Cpar}{3} \nn\\
\alphaj &=& \frac{\Cpol R_j F_j}{\Cp}. 
\eea
Note that $\alphaj$ does not vary with position, since we assume that $R_j$ and $F_j$ are
constants for each grain species.  The contribution of grain species $j$ to the intensity 
$I$ is given by
\be
I_j=\Cp \Bnu c_j \left(\Sigma-\alphaj \Sigma_2\right),
\label{eq:I}
\ee
where $\Sigma$ is the surface density and $\Sigma_2$ is a related quantity defined by
\bea
\Sigma &=& \int \rho ~ds \nn\\
\Sigma_2 &=& \int \rho \left( \frac{\cos^2\gamma}{2}-\frac{1}{3} \right) ~ds.
\eea
Note that $\Sigma$ and $\Sigma_2$ depend on the density and magnetic structure of the cloud, 
but not the grain properties.

The polarization percentage is defined by
\be
p=\frac{\sqrt{Q^2+U^2}}{I},
\label{eq:pdef}
\ee
where $Q$, $U$, and $I$ are obtained by summing equations \ref{eq:Q}, \ref{eq:U}, and \ref{eq:I} over
grain species $j$.  It is easy to show that equation \ref{eq:pdef} becomes
\be
p=\meanalpha \frac{\sqrt{q^2+u^2}}{\Sigma-\meanalpha\Sigma_2},
\label{eq:p}
\ee
where $\meanalpha$ is a weighted mean of $\alphaj$ defined as follows:
\be
\meanalpha=\frac{\Sigma_j ~ \alphaj\Cp\Bnu c_j}{\Sigma_j \Cp\Bnu c_j}
\label{eq:meanalphadef}
\ee

Equations \ref{eq:q}, \ref{eq:u}, and \ref{eq:p} show that the polarization pattern is determined by
the density and magnetic structure of the gas, plus a single parameter $\meanalpha$
related to the grain cross-sections and alignment properties.  We estimate $\meanalpha$ as follows.
We assume that the optimal magnetic field geometry, in which $\gamma=0$ and $\psi=const$,
gives rise to the maximum polarization percentage $p_{max}$ that is normally observed.  
The maximum polarization percentage is obtained from equation \ref{eq:p}, with the help of 
equations $\ref{eq:q}$ and \ref{eq:u}:
\be
p_{max}=\frac{\meanalpha}{1-\meanalpha/6}.
\label{eq:pmax}
\ee
Since the maximum polarization percentage observed in sub-millimetre polarization maps is
rarely greater than $10\%$, equation \ref{eq:pmax} suggests that $0.1$ is a reasonable choice 
for $\meanalpha$.   Using this choice, our approach predicts the magnitude of the polarization 
percentage obtained from our models, which is generally less than $p_{max}$.  
This is in contrast to WK90 who work with normalized quantities.  
We caution the reader that $\meanalpha$ could, in principle, vary from region to region, which would 
affect the magnitude of the polarization percentage that we predict.  It would also slightly affect the shapes of the 
polarization profiles, since $\meanalpha$ apppears in the denominator of equation \ref{eq:p} and cannot be factored out.

\section{Results}
\label{sec:results}
\subsection{Transverse Fields Across Filamentary Clouds?}
\label{sec:transverse}
Polarimetry only maps the component of the magnetic field parallel to the plane of the sky.
Suppose that the magnetic field threading the integral-shaped filament in Orion A contains only this 
plane-of-sky component.
If this were true, the Matthews and Wilson (2000) map, as well as the smaller scale map by Schleuning (1998),
would suggest that we are seeing either a filament or an edge-on sheet being
impaled by a well-ordered field that is perpendicular to the filament axis or the midplane of the sheet.
While this might explain the overall sense of the polarization vectors, it cannot easily account for the depolarization
of these maps seen toward the centre of the putative sheet or filament.
For such a transverse field model, it is easy to verify from equations \ref{eq:q}, \ref{eq:u}, and \ref{eq:p} that $p$
would be constant over the entire filament, since $\zeta=0$ everywhere and $\psi=const$.
The analysis of Section \ref{sec:analysis}
would not predict the observed depolarization along the axis of the filament or the midplane of
the sheet.  It could, however, be accomplished if the polarization were due to grains that are preferentially more
spherical or poorly aligned in dense regions.

We show in Section \ref{sec:patterns} that helical fields threading a filamentary cloud can result in depolarization toward
the filament axis, otherwise similar in appearance to what would be expected from the above transverse field scenario.
However, the depolarization is due to the 3-dimensional structure of the field in this case and does not depend 
on the grain shapes or their alignment.
 
\subsection{Polarization Patterns and Depolarization Effects for Helical Fields}
\label{sec:patterns}
We present polarization maps of our helically magnetized models of filamentary clouds
using the method discussed in Section \ref{sec:analysis}.  We first restrict our parameter space 
by only showing results for filaments whose axes lie in the plane of the sky.  In Section \ref{sec:inc}, we 
show maps for a filament at several inclination angles on the sky.

\begin{figure}[ht]
\psfig{file=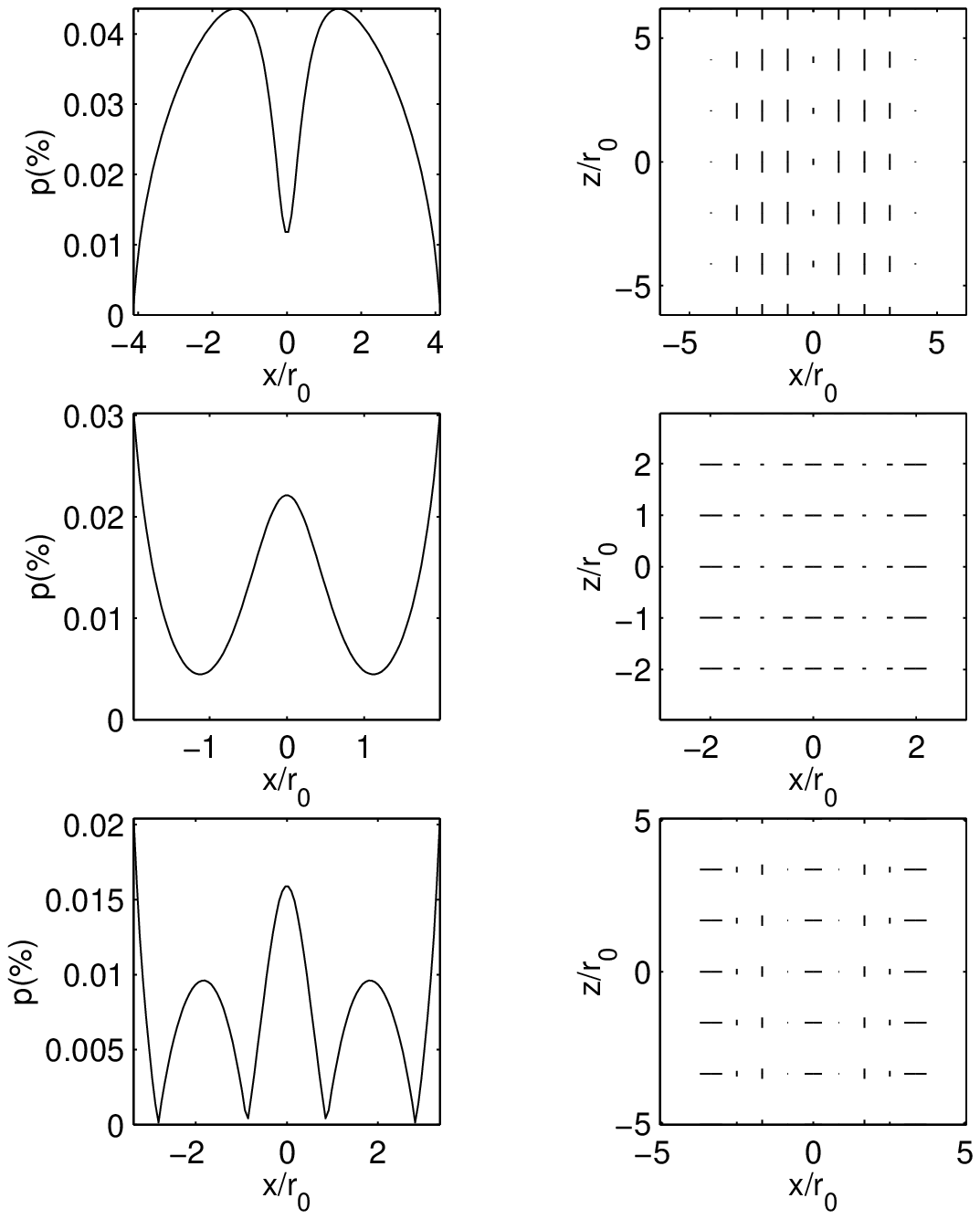,width=\linewidth}
\caption{We show examples of the three types of polarization maps that we obtain for filaments oriented perpendicular
to the line of sight.  ({\bf Top row:}) Type 1: A toroidally dominated field pattern ($C=0.62$, $\Gz=5.1$, $\Gphi=13.6$).
({\bf Centre row:}) Type 2: A poloidally dominated field pattern ($C=0.31$, $\Gz=8.7$, $\Gphi=11.9$).
({\bf Bottom row:}) Type 3: A mixed field pattern ($C=0.54$, $\Gz=10.4$, $\Gphi=11.4$).
}
\label{fig:types}
\end{figure}

Figure \ref{fig:types} shows three representative models to illustrate the general types of behaviour that we find in our maps.  
We refer to these qualitative patterns as types 1 to 3 for simplicity, although we emphasize that the underlying
models represent a continuum in parameter space (see FP1).
In panel a) we show a model (type 1) where the polarization vectors are everywhere parallel to the filament.  
The most striking feature of the map is the depolarization along the axis.  Qualitatively, this map shares some features with 
the Matthews and Wilson (2000) map of the Orion filament, specifically the overall orientation of the polarization vectors and
the depolarization toward the central regions.
Panel b) shows a model (type 2) whose polarization vectors are oriented opposite to the type 1 model in panel a).
If such a pattern were observed, it could be misinterpreted as the result of a purely poloidal field.  These models 
have two depolarized regions, with one on either side of the filament axis.  The polarization on the filament axis is a
local maximum, although the strongest polarization in the map is usually at the surface of the filament. 
Panel c) shows a more complicated case (type 3), where there are $90^\circ$ flips in the orientation of the polarization
vectors.  The polarization percentage passes through zero at all locations where a flip
in orientation occurs.  We classify any patterns containing such flips in orientation as type 3.

It may seem surprising that filamentary clouds that are threaded by helical fields always have polarization vectors that are aligned
parallel or perpendicular to the filament.  This result is obtained regardless of whether the filament is parallel to the plane
of the sky or inclined at at some angle.  It is easy to understand this by the following argument (see Carlqvist \& Kristen (1997)
and FP1).  Any line of sight passing through a filament encounters a given magnetic flux tube twice; first on the front side of the filament, 
and then on the back.  By symmetry, the plane-of-sky components of the magnetic field vectors are mirror images of each other at 
these positions, so that $\psi$ is an odd function of the position $s$ along each line of sight.  Therefore, equations \ref{eq:q} and 
\ref{eq:u} show that the contributions to $u$ exactly cancel, while the contributions to $q$ add in equal proportion.

From the definition of the polarization angle (equations \ref{eq:angle1} and \ref{eq:angle2}), we observe that the polarization 
vectors are oriented perpendicular to the filament ($\chi=0$) when $q>0$ and parallel to the filament ($\chi=\pi/2$) when $q<0$.  
The dominant magnetic field component on each line of sight determines the orientation of the field vectors.  This is easily seen by the 
following argument.  Consider first a purely poloidal field threading a filament oriented in the North-South direction,
parallel to the plane of the sky.  The Stokes vector $q>0$ in this case, so that the polarization vectors are perpendicular to the 
filament, since $\cos 2\psi=1$ everywhere.  On the other hand, if the field were purely toroidal,  
$\cos 2\psi=-1$, so that $q$ is negative and the polarization vectors are parallel to the filament.  
Now consider the case of a helical field threading a filament oriented parallel to
the plane of the sky.  It is obvious that the poloidal and toroidal components of the magnetic field provide competing 
contributions to $q$ through $\psi$, and that the polarization direction at each position is determined by the field component that makes
the dominant contribution to equation \ref{eq:q} along each line of sight.

The competing contributions to $q$ from the poloidal and toroidal field components may partially or completely cancel 
at some positions, whenever the poloidal and toroidal fields are of comparable strengths.  
This is the explanation for the depolarized regions seen in all three panels of Figure \ref{fig:types}.  

\subsection{Exploration of the Parameter Space}
We explore our parameter space in the same spirit as we did in FP1, FP2, and Fiege \& Pudritz 2000c (Hereafter FP3).  
We compute polarization maps for models 
scattered randomly throughout our 3-dimensional parameter space ($C$, $\Gz$, $\Gphi$), subject to the observational constraints
discussed in FP1.  A slight complication is that we exclude some models that were found to be highly unstable
to short wavelength, rapidly growing MHD instabilities in FP2.  We note that a similar criterion was used in 
FP3 to select ``parent'' filaments for our prolate core models.  Specifically, we exclude models for which
$-\omega_{max}^2/(4\pi G\rho_c) > 0.05$,
where $-\omega_{max}^2$ is the maximum squared growth rate, and $\rho_c$ is the central density of the filament (See FP2, Figure 13).

\begin{figure}[ht]
\psfig{file=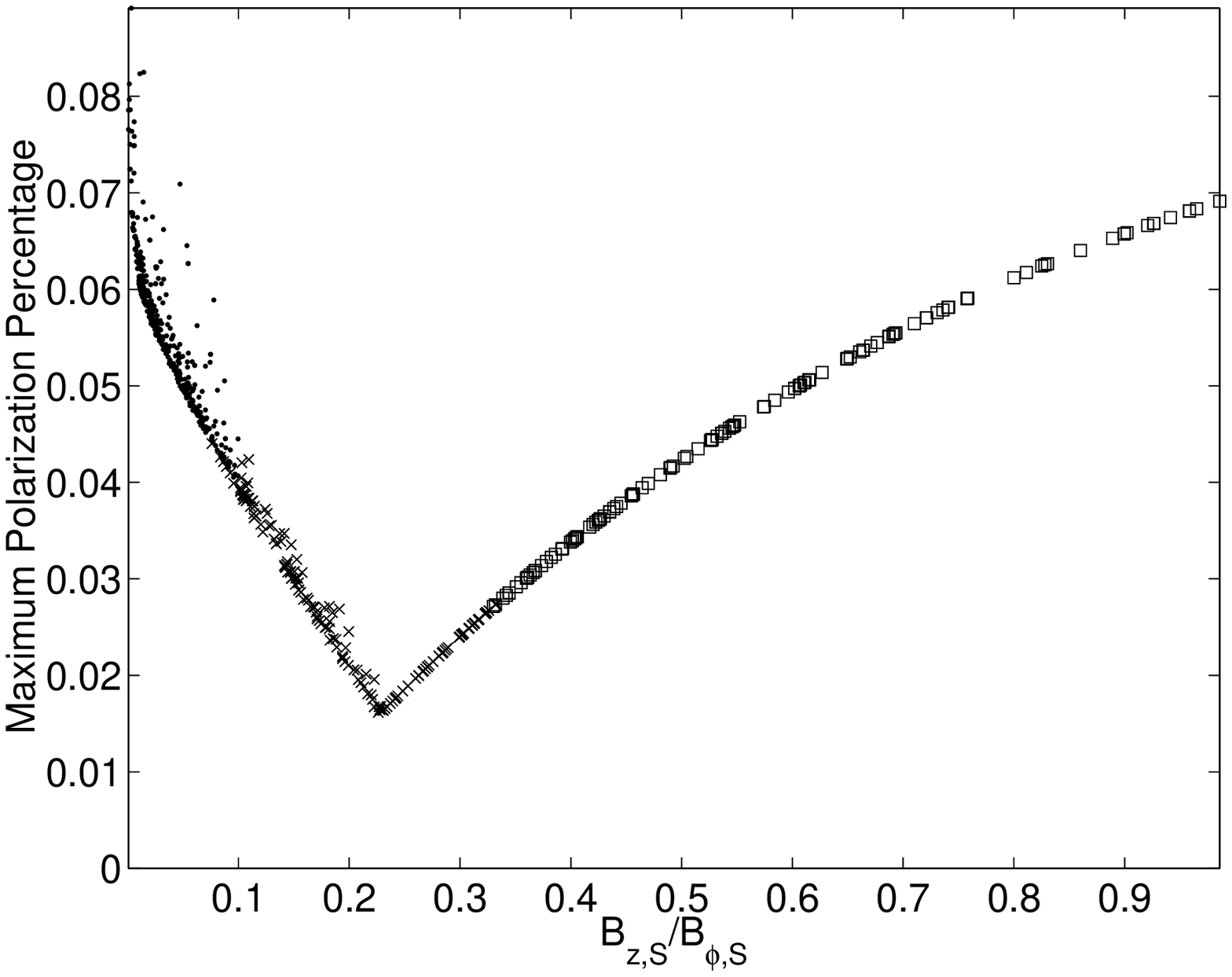,width=\linewidth}
\caption{We show the polarization percentage as a function of $\ratio$.  According to the classification scheme discussed in Section
\ref{sec:patterns}, dots represent type 1 patterns, squares represent type 2 patterns, and x's represent type 3 patterns.}
\label{fig:space}
\end{figure}

In Figure \ref{fig:space}, we show how the polarization pattern depends on $\ratio$, where $\Bzs$ and $\Bphis$ are respectively 
the poloidal and toroidal field components, evaluated at the surface of the filament:
\be
\frac{\Bzs}{\Bphis}=\frac{ e^{-C} \Gz}{\Gphi}.
\ee
The dots, squares, and x's represent models of types 1, 2, and 3 respectively.
Note that $\ratio$ is small for many of our models because the toroidal magnetic field dominates in the outer envelope, near the radius of
pressure truncation.  The ratio of $\Bz/\Bphi$ is substantially higher in the interior regions of the filament.

An interesting feature of Figure \ref{fig:space} is that the polarization patterns are of the first type for a large portion of our
models.  Thus, many of our models are qualitatively similar to the Matthews and Wilson (2000) map of the Orion filament, discussed 
in Section \ref{sec:patterns}.  We find this type of pattern for most models with $\ratio\appleq 0.1$.  Generally,
polarization patterns of the second type occur when $\ratio\appgeq 0.33$, and the third type occurs for intermediate
values between about 0.1 and 0.33.  This may be understood as follows.  When $\ratio$ is small, the models are dominated by the
toroidal field component so that the poloidal field is ineffective at canceling the polarization due to the toroidal field,
as discussed in Section \ref{sec:patterns}.  Thus, the polarization pattern is of type 1.  
However, as $\Bzs$ is increased relative to $\Bphis$, the competition becomes stronger until $\Bzs$ becomes dominant along 
some line of sight, which first occurs when $\ratio \approx 0.1$.  This results in a $90^\circ$ flip in the polarization vectors at this
position, which we categorize as the first Type 3 model.  The exact opposite occurs when $\ratio$ is increased past about $0.33$.   
Past this point, the toroidal field component becomes too weak compared
to the poloidal field to produce a flip in the polarization vectors resulting in Type 2 patterns.

\begin{figure}[ht]
\begin{minipage}{\linewidth}
\psfig{file=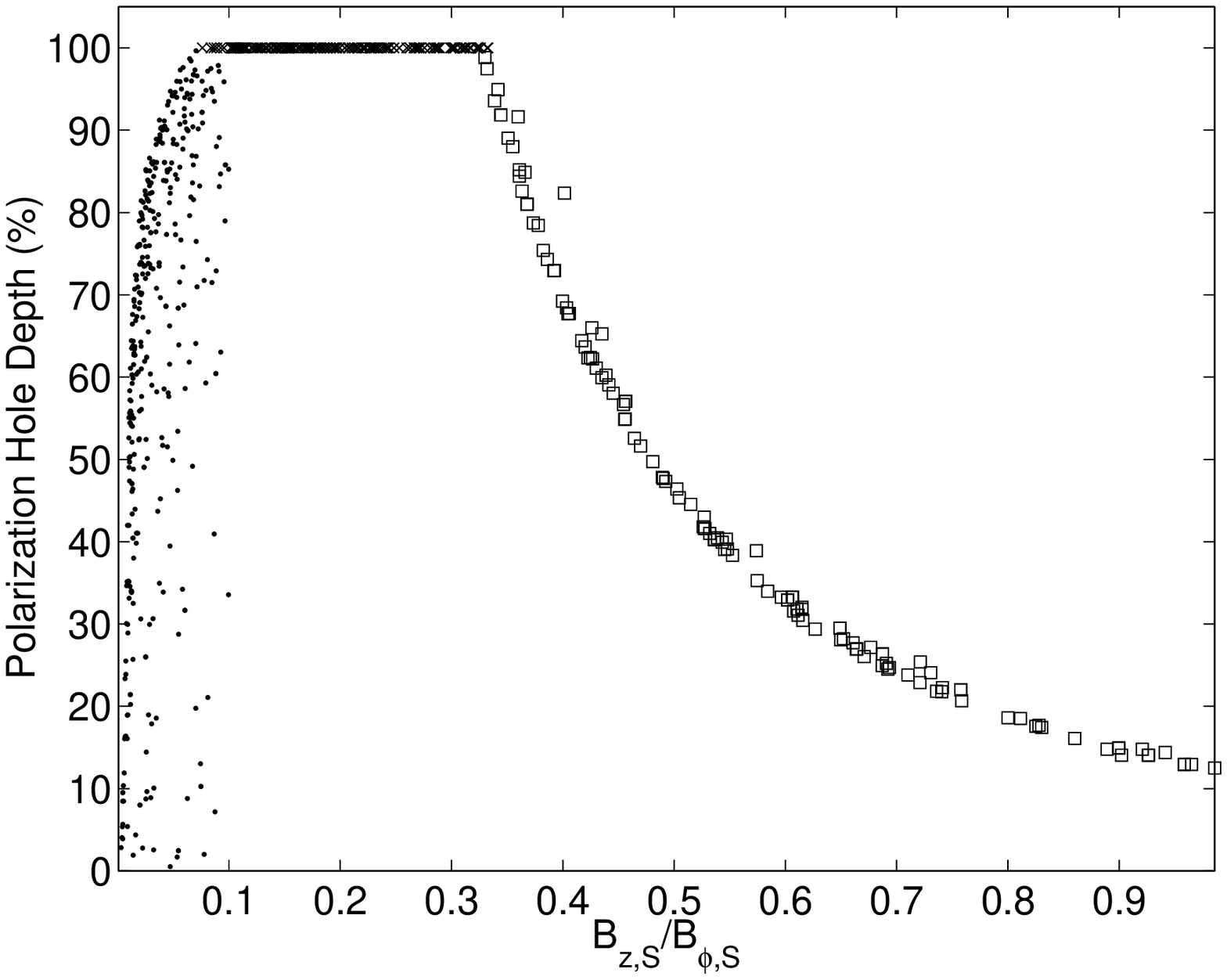,width=0.49\linewidth}
\psfig{file=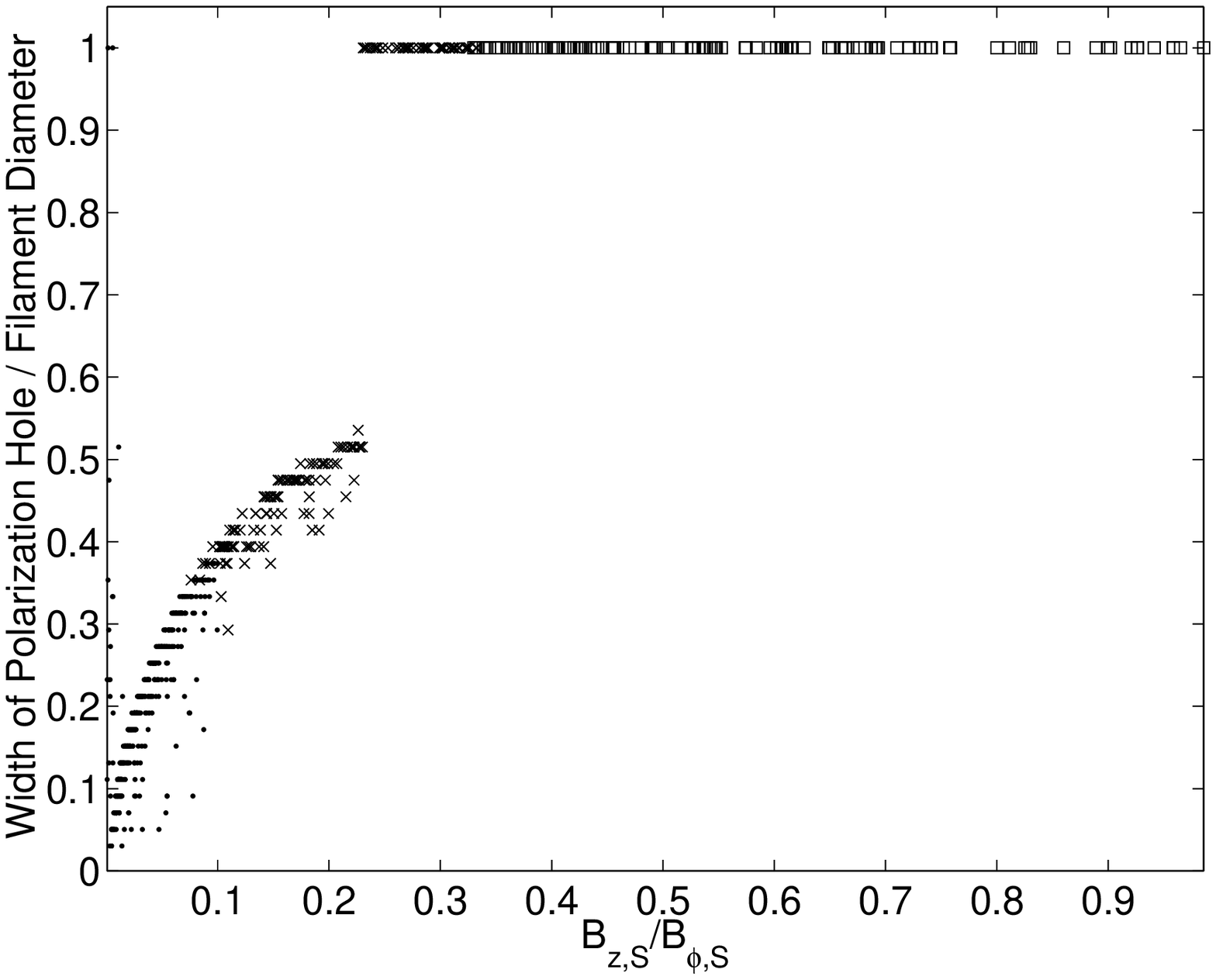,width=0.49\linewidth}
\end{minipage}
\caption{We show the relative depolarization of our models (left panel), and the width of the depolarized region (right panel).
The dots, squares, and x's use the same classification scheme as in Figure \ref{fig:space}.}
\label{fig:depolarized}
\end{figure}

Figure \ref{fig:depolarized} shows the degree to which the emission is depolarized for all three types of polarization pattern.
We define the polarization hole depth as $(P_{max}-P_{min})/P_{max}$, where $P_{max}$ is the maximum polarization percentage in the map and
$P_{min}$ is the local minimum polarization percentage at the location of the polarization hole with the lowest polarization.  
We find that the polarization hole depth generally increases, with scatter, as a function of $\ratio$ for Type 1 models from $0\%$ when 
$\ratio=0$ to $100\%$ when $\ratio=0.1$.  This is easily understood by essentially the same argument given in the previous paragraph.
There is no significant polarization hole when $\ratio$ is small because contributions to the polarization arising from 
the poloidal field do not effectively cancel the larger contributions from the toroidal field along any line of sight.  
The depth of the polarization hole increases until $\ratio \approx 0.1$,
where the first Type 3 pattern emerges.  Note that the polarization hole depth is always $100\%$ for Type 3 models, since $P_{min}=0$ at
the locations where the oriention of the polarization vectors flips by $90^\circ$.  Increasing $\ratio$ decreases the depth of the 
polarization hole for Type 2 models, since the contribution to the polarization from the toroidal field becomes progressively less effective at 
canceling the polarization due to the dominant poloidal field.

We define the width of the depolarized region as the distance between the polarization maxima on either side of the polarization hole.
Panel \ref{fig:depolarized}b shows the ratio of this width divided by the filament diameter.  
Generally, this ratio increases with $\ratio$ for Type 1 models, from nearly $0$ to about $0.4$,
which occurs at the point where the models change to Type 3.  This fractional width continues to increase past this point until $\ratio \approx 0.23$, where
the polarization width jumps discontinuously to the full width of the filament.  This happens because of a change in the qualitative behaviour of the 
polarization vectors near the edge of the filament.  Note that the polarization percentage shown in Figure \ref{fig:types} is a minimum at the outer edge of the filament
for Type 1 models, but a maximum for Type 2 models.  As we move through a sequence of Type 3 models, the polarization at the edge gradually 
increases as a function of $\ratio$ until $\ratio \approx 0.23$.  Past this point, the maximum polarization for the filament occurs at the edge, so the width
of the depolarized region becomes equal to the width of the filament.

\subsection{The Effect of Inclination Angle}
\label{sec:inc}
\begin{figure}[ht]
\psfig{file=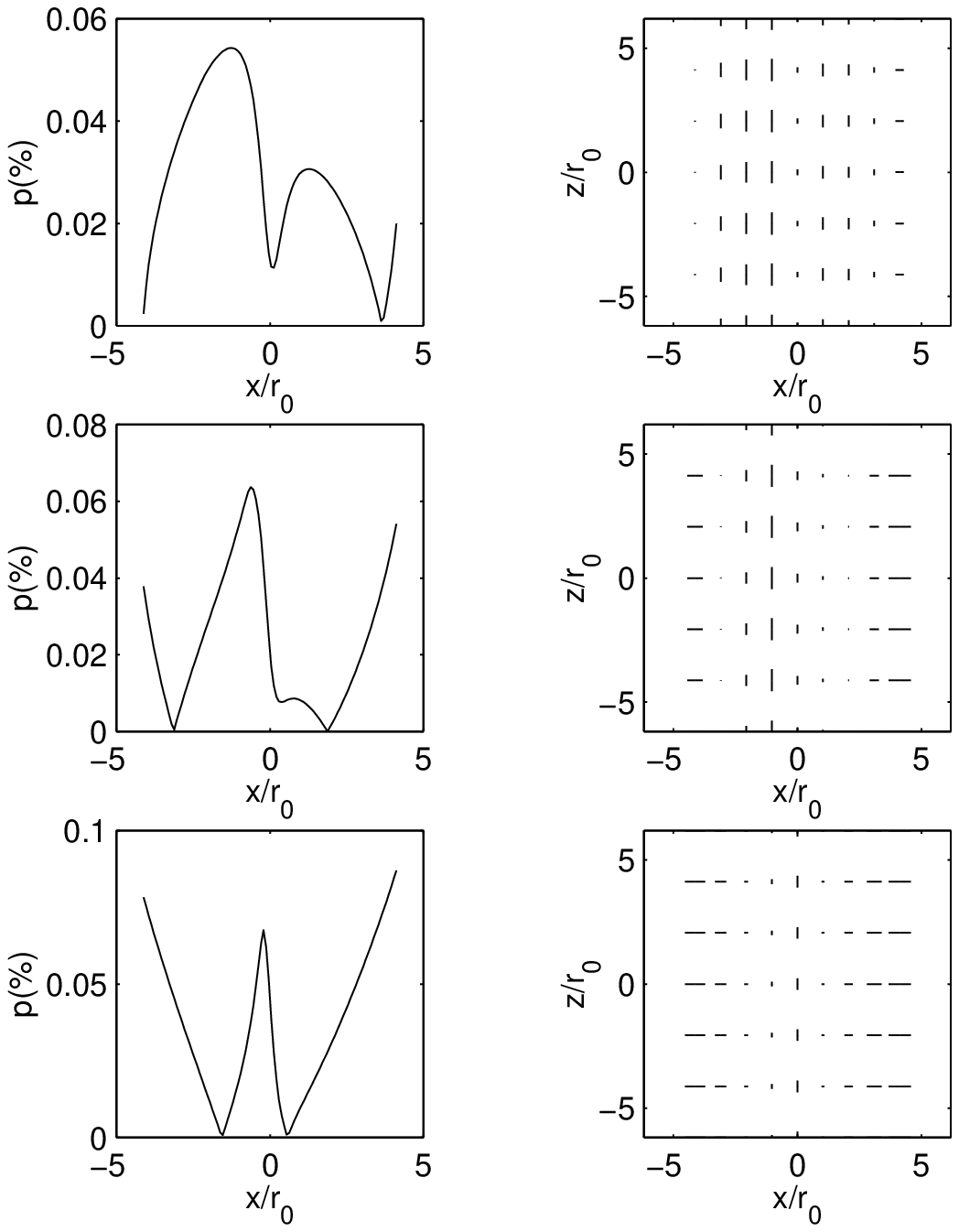,width=\linewidth}
\caption{We show the polarization patterns expected for one of our models oriented at $10^\circ$ (top row), 
$30^\circ$ (centre row), and $60^\circ$ (bottom row) relative to the plane of the sky.
The model used here is the same one used in the top row of Figure \ref{fig:types}.}
\label{fig:inc}
\end{figure}

The polarization patterns due to helically magnetized filaments become more complicated for filaments that are inclined relative to 
the plane of the sky.  As an example, in Figure \ref{fig:inc}, 
we show maps of one filament tilted $10^\circ$, $30^\circ$, and $60^\circ$ relative to the plane of the 
sky.  The filament model shown in Figure \ref{fig:inc} is the same one as shown in the top two panels of Figure
\ref{fig:types}.  Each polarization vector always remains parallel or perpendicular to the symmetry axis for the reasons
discussed in Section \ref{sec:patterns} above, but the patterns become asymmetric relative to the central axis of the filament.  
It is particularly notable that filaments inclined relative to the line of sight usually result in asymmetric 
polarization patterns that include a $90^\circ$ flip in the orientation of the polarization vectors on one or both sides of the filament. 

\section{Discussion \& Summary}
The main purpose of this paper is to calculate polarization maps for the models of filamentary molecular clouds that we presented in FP1 and FP2.
When the filament is oriented parallel to the plane of the sky, the polarization patterns resulting from our models can be 
qualitatively classified into three main types of behaviour.  The polarization vectors are parallel to the symmetry axis of the filament 
(Type 1) when the toroidal field is dominant, and perpendicular (Type 2) for models whose poloidal field dominates. 
The third type of pattern is more complicated, with polarization vectors that flip from being aligned parallel to the filament axis to
perpendicular at some radius or radii (in projection).
Generally, the first type of pattern occurs when $\ratio \appleq 0.1$, while the second type
occurs $\ratio\appgeq 0.33$.  The third type of polarization pattern occurs at intermediate values of $\ratio$
Our most important result is that helical fields result in ``polarization holes,'' in which the emission is
depolarized at some positions in the interior of the filament.  Our Type 1 models qualitatively agree with the 
polarization structure of the ``integral-shaped'' filament in Orion A (Matthews \& Wilson 2000).

Many of our models result in polarization patterns that contain $90^\circ$ flips in the orientation of the polarization vectors.
The flips are symmetric about the symmetry axis when a filament with comparable poloidal and toroidal magnetic field
strengths is oriented parallel to the plane of the sky.  Filaments that are inclined relative to the line of sight generally result in 
asymmetric polarization flips on one or both sides of the central axis of the filament.

We have considered only models of non-fragmented, cylindrically symmetric
filaments in this paper.  This is obviously an idealization of real
filamentary clouds, many of which have suffered gravitational
fragmentation and formed strings of embedded cores (cf. Schneider \&
Elmegreen 1979, Dutrey et al. 1991; see also FP2 and FP3), and may  
also  deviate from perfect cylindrical symmetry.  We will address the polarization
patterns of embedded cores in a future analysis.  It is difficult to
comment on the general effects of asymmetry.
However, it is worthwhile to consider the simplest type of
non-axisymmetric perturbation, that corresponding to a kink mode in which the filament is
bent into a transverse sinusoidal wave.  If the sinusoid lies  
in the plane of the sky, the front to back symmetry is preserved and the polarization
vectors remain locally parallel or perpendicular to the filament.  The
depolarization along the central axis of the filament relies on a
cancellation between contributions to the polarization from the central
``backbone'' of poloidal flux and the surrounding envelope dominated by
the toroidal field component.  Since these features would be preserved in
such a perturbation, the polarization hole would also likely remain.
If, on the other hand, the sinusoidal perturbation is perpendicular to
the plane of the sky, then the front to back symmetry is broken
and the polarization vectors would no longer be strictly parallel
or perpendicular to the filament.  A segment which is bent away from the observer
is more compressed on the front side and more rarified
on the back, so the front side would contribute more to the polarization.
Assuming that the filament is oriented in the  north-south direction and 
the field lines have right-handed (positive) helicity, the polarization  
vectors would shift toward the north-east/south-west plane if they were  
originally oriented in the north-south direction.  The polarization hole 
would be preserved for the same reasons as in the case of a perturbation 
parallel to the plane of the sky.  Thus, we see that non-axisymmetric distortions 
of a filament containing a helical field still ought to show a ``polarization hole''.

Ignoring systematic variations of grain shapes and alignment efficiencies with density, 
we have shown that the ``polarization hole'' effect has a very simple
and entirely geometric explanation if the field lines are twisted into helices.  
It might alternately be explained if the emitting dust grains are poorly aligned in the central 
regions of filamentary clouds, but this requires a greater understanding of dust grain physics.

More rigorous tests of our models will require a full set of polarization maps of filamentary clouds.   
Such a data set, when combined with well determined radial density profiles, will provide an 
excellent challenge to all cloud models, not only our own.
{\em Readers wishing to use our models to compare with data should contact J. Fiege to obtain fits of our model to data,
or polarization maps for any inclination angle and set of input parameters ($C$, $\Gz$, and $\Gphi$).}
It is probably true that even with the density profiles and polarization structure of filamentary clouds 
determined by observations, it would still be necessary to make spatially resolved Zeeman measurements
in order to conclusively prove the existence of helical fields.  By making such maps, we might one day
decide whether grain properties or the structure of the magnetic field is responsible for the polarization hole effect.   

\section{Acknowledgements}
The authors thank B.C. Matthews and the anonymous referee for useful comments on a draft of the manuscript.
J.D.F. acknowledges the financial support of McMaster University, CITA, and the Natural Sciences and Engineering Research
Council of Canada (NSERC) during this research.  The research of R.E.P. is supported by a grant from NSERC.

\end{document}